\documentclass[graybox]{svmult}

\usepackage{mathptmx}       
\usepackage{helvet}         
\usepackage{courier}        
\usepackage{type1cm}        
\usepackage{makeidx}         
\usepackage{graphicx}        
\usepackage{multicol}        
\usepackage[bottom]{footmisc}

\makeindex             

\begin{document}

\title*{A Streamline Selection Technique Overlaying with Isosurfaces}
\author{Shiho Furuya, Takayuki Itoh}
\institute{Shiho Furuya \at Ochanomizu University, \email{shiho@itolab.is.ocha.ac.jp}
\and Takayuki Itoh \at Ochanomizu University \email{itot@is.ocha.ac.jp}
\and (This paper is presented at TopoInVis 2017, 2/27/2027 at Keio University, Japan. Low resolution images are used in this PDF file due to file size limitation of arXiV. For better images, please see http://itolab.is.ocha.ac.jp/$\sim$itot/paper/ItotRICPE109.pdf)
}
%
%
\maketitle

\abstract{
Integration of scalar and vector visualization has been an interesting topic. This paper presents a technique to appropriately select and display multiple streamlines while overlaying with isosurfaces, aiming an integrated scalar and vector field visualization. The technique visualizes a scalar field by multiple semitransparent isosurfaces, and a vector field by multiple streamlines, while the technique adequately selects the streamlines considering reduction of cluttering among the isosurfaces and streamlines. The technique first selects and renders isosurfaces, and then generates large number of streamlines from randomly selected seed points. The technique evaluates each of the streamlines according to their shapes on a 2D display space, distances to critical points of the given vector fields, and occlusion by isosurfaces. It then selects the specified number of highly evaluated streamlines. As a result, we can visualize both scalar and vector fields as a set of view-independently selected isosurfaces and view-dependently selected streamlines.
}

\section{Introduction}

Scalar and vector visualization techniques have been almost independently evolved in the research field of volume visualization.
Meanwhile, it often happens that we want to simultaneously and comparatively visualize both scalar and vector fields in a same physical space.
Cluttering is a serious problem while visualizing both scalar and vector fields in a single display space.
This paper presents a technique which simultaneously visualizes both scalar and vector fields in a same physical space.
We visualize a scalar field as a set of semitransparent isosurfaces, and a vector field as a set of streamlines.
The presented technique features automatic selection of specified number of streamlines from the following three aspects: geometric meaningfulness evaluated by information entropy, distance from critical points, and occlusion by isosurfaces.

This paper is an extended version of our previous poster presentation \cite{Furuya2008}, which additionally introduces subjective evaluations, and a case study with a weather simulation dataset which visualized relationship between an air flow field and high/low pressure regions.

\section{Related Work}

\subsection{Streamline selection}

Automatic streamline selection techniques have been firstly developed for 2D flow visualization, and then extended to 3D visualization.

Mebarki et al. \cite{Mebarki2005} claimed that uniformly placed sets of streamlines can be generated by specifying positions of seed points of streamlines as most distant from previously generated streamlines.
Li et al. \cite{Li2008} presented a 2D visualization technique which generates appropriately distant sets of streamlines while controlling the number of streamlines.

Many 3D streamline selection techniques have been recently presented.
Ye et al. \cite{Ye2005} presented a technique which prepares several templates to specify positions of seed points of streamlines around critical points of 3D flow fields, so that users can appropriately generate streamlines which represent the features of flow fields.
Li et al. \cite{Li2007} presented a streamline generation technique which sets seed points on a plane in a 3D space, so that we can avoid cluttering of streamlines.
Marchesin et al. \cite{Marchesin2010} extended our idea \cite{Furuya2008} for view-dependent streamline removal and addition. 
Several recent studies \cite{Lee2011} \cite{Tao2013} estimate viewpoints while selecting streamlines in 3D flow fields, so that we can observe the flow fields from the best viewpoint which brings comprehensive visualization results.

However, most of existing streamline selection techniques just attempt to optimize the visualization of flow fields.
On the contrary, our technique presented in this paper selects streamlines for simultaneous visualization of scalar and vector fields in a same physical 3D space.

\subsection{Integrated visualization of scalar and vector fields}

We often have 3D flow dynamics simulation datasets which contain both scalar and vector fields in a same physical space.
It is important to simultaneously visualize both scalar and vector fields to understand the flow dynamics.
Direct volume rendering is a convenient method for this purpose.
It is not usually difficult to control transfer functions of volume rendering to overlay indirect vector field visualization such as arrow plots and streamlines.
Such representation has been applied since the early days of visualization research.
Crawfis et al. \cite{Roger1993} represented scalar and vector fields by using splatting-based volume rendering and particle animations.
Hong et al. \cite{Hong1995} applied glyphs which represent vector fields while overlaying with volume rendering for scalar fields.

On the other hand, it sometimes happens that indirect volume rendering is more preferable.
We may have a situation that visualization results should be stored as indirect datasets (e.g. isosurfaces or iso-contours for scalar fields, streamlines or arrow plots for vector fields).
Or, runtime environments which GPU programming is not available are still existing.
Combination of indirect visualization techniques for scalar and vector fields is another solution for these situations, as we apply isosurfaces and streamlines in this paper.
Actually, there are many studies which applies isosurfaces and streamlines for integrated scalar and vector field visualization \cite{Merhof2009} \cite{Sch2012}.
However, these studies did not apply automatic streamline selection techniques.

It is also meaningful to apply isosurfaces to represent flow fields \cite{Edmunds2012}.
For example, Laramer et al. \cite{Laramee2004} applied velocity isosurfaces with streamlines to effectively represent flow fields.

\section{Presented visualization technique}

This section describes the processing flow and technical detail of the presented technique.  Following is the brief processing flow of the technique.

\begin{description}
\item[Step 1:] Extract critical points of given scalar and vector fields.
\item[Step 2:] Generate semitransparent isosurfaces to represent a scalar field.
\item[Step 3:] Temporality generate large number of streamlines.
\item[Step 4:] Select specified number of streamlines which are highly evaluated based on information entropy, distances to critical points, and occlusion by isosurfaces.
\end{description}

The following subsections describe the technical detail of the each process of the presented technique.

\subsection{Critical point extraction} \label{sec:cp}

We often intensively visualize around critical points of scalar or vector fields to discover interesting phenomena.
In other words, we can often visualize interesting phenomena by generating isosurfaces and streamlines around critical points.
For example, in case of visualizing weather simulation results, we can often discover interesting scalar fields around high (or low) pressure regions, or interesting vector fields around centers of vortices.

Accordingly, we developed a critical point extraction technique for scalar and vector fields, to intensively visualize the scalar and vector values around the critical points.
We apply a critical point extraction technique for scalar fields presented by Itoh et al. \cite{Itoh1995}.
This technique extracts local maximum or minimum points of scalar fields by comparing scalar values of grid-points connected by element-edges.

We also apply a critical point extraction technique for vector fields presented by Koyamada et al. \cite{Koyamada1998}.
This technique determines that a position $P(p,q,r)$ is a critical point of a vector field inside a tetrahedral cell, if $P$ satisfies the following equation:
\begin{eqnarray}
\left(
\begin{array}{c}
 p \\
 q \\
 r \\
\end{array}
\right)
= M_v^{-1} 
\left(
\begin{array}{c}
 -u_3 \\
 -v_3 \\
 -w_3 \\
\end{array}
\right)
, \hspace*{5mm}
M_v= 
\begin{array}{|ccc|}
u_0 - u_3 & u_1 - u_3 & u_2 - u_3 \\
v_0 - v_3 & v_1 - v_3 & v_2 - v_3 \\
w_0 - w_3 & w_1 - w_3 & w_2 - w_3 \\
\end{array} \nonumber
\\
0 \le  p \le 1, 0 \le q \le 1, 0 \le r \le 1 \nonumber
\end{eqnarray}
Here, $(u_i,v_i,w_i)$ is a vector value at the $i$-th grid-point of the current tetrahedral cell.
Also, we suppose the current tetrahedral cell can be divided into four sub-tetrahedral regions by connecting $P$ and each of vertices of the tetrahedral cell.
Ratios of volumes of the sub-tetrahedral regions are $p$, $q$, $r$, and $(1-p-q-r)$.

\subsection{Visualization of scalar fields} \label{sec:scalar}
This technique uses the critical points of a scalar field while generating isosurfaces.
Our implementation lists scalar values at the extracted critical points and dispalys as a user interface widget.
When a user selects one of the displayed scalar values, the technique generates an isosurface by applying Marching Cubes.
Our implementation allows to display semi-transparent isosurfaces while selecting multiple scalar values.

Our current implementation requires manual adjustment the opacity of isosurfaces. 
As a future work, we would like to automatically adjust by applying automatic transfer function design techniques \cite{Kindlmann1998}\cite{Takahashi2004}.

\subsection{Visualization of vector fields} \label{sec:vector}
\subsubsection{Information entropy for streamline selection} \label{sec:ent}
This section describes an automatic streamline selection technique applying information entropy.
First, this technique uniformly generates large number of streamlines thoroughly inside an input volume region.
Then, the technique calculates information entropy for each of the streamlines applying a technique presented by Takahashi et al. \cite{Takahashi2005}.
We automatically select a specified number of streamlines which have larger information entropy values.
Information entropy has been well applied to variety of recent visualization techniques \cite{Xu2010}.
The following is an equation applied in our technique to calculate information entropy $E$:
\begin{equation}
E = - \frac{1}{ \log_{2} (1+m)} \sum^m_{j=0}\frac{d_{j}}{l_j}\log_{2}\frac{d_{j}}{l_j} \label{eq:entropy}
\end{equation} 
Here, we suppose streamlines consist of short segments.
$l_j$ and $d_j$ are the length of the $j$-th segment in the 3D physical and 2D display spaces respectively, and $m$ is the number of the segments in a streamline.

We can reselect appropriate sets of streamlines by calculating $E$ for all of the pre-generated streamlines, when the viewpoint is moved or view direction is modified.
Our implementation supports drag operations of pointing devices to control the viewpoint and view direction, and execute the automatic streamline selection when the drag operations are finished.

\begin{figure}[ht]
  \begin{center}
   \includegraphics[width=100mm]{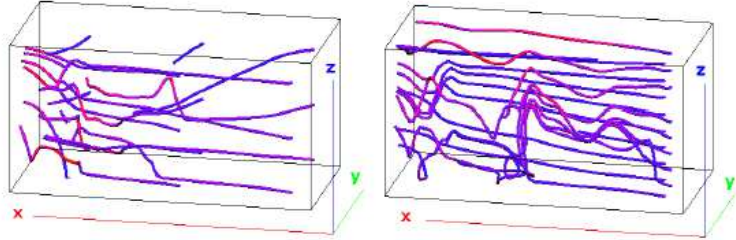}
  \end{center}
  \caption{\small Example of streamline generation.
(Left) Uniformly generated streamlines.
(Right) Streamlines automatically selected based on information entropy.
}
  \label{fig:1-2}
\end{figure}

Figure \ref{fig:1-2}(Left) shows an example of uniformly generated streamlines, and Figure \ref{fig:1-2}(Right) shows an example of streamlines selected based on information entropy, where both examples draws 18 streamlines.
Our technique successfully represents the vector field comparing with uniformly generating streamlines.

Furthermore, our streamline selection technique takes into account critical points, occlusion by isosurfaces, and density of streamlines, as described in below sections.

\subsubsection{Streamline selection with critical poins}\label{sec:cpp} 
Often we would like to visualize the vector fields around critical points because interesting phenomena are often found there \cite{Koyamada1998}.
However, we experimentally found that $d_j$ in equation (\ref{eq:entropy}) of streamlines around critical points tends to be small and therefore such streamlines are not often selected while simply applying the information entropy. 
Therefore, our implementation optionally selects at least one streamline generated from vicinity of each of critical points.

\subsubsection{Occlusion by isosurfaces}
We would like to avoid un-preferable visualization results that many streamlines are occluded by isosurfaces.
Our technique revises the information entropy by the following equation to preferentially select un-occluded streamlines:
\begin{equation}
E' = (1 - \frac{m_{0}}{ m })E + (\frac{m_{0}}{m})(1-\alpha ) E \label{eq:re-en},
\end{equation} 
where $m_{0}$ is the number of occluded segments of a streamline, and $\alpha$ is the opacity of an isosurface.

Our implementation displays multiple isosurfaces as described in Section \ref{sec:scalar}.
Supposing the $i$-th segment ($1\leq i \leq m$) of a streamline is occluded by the $k$-th isosurface which is rendered with the opacity value $\alpha_{k}$, we define the weight of the $i$-th segment as follows:
\begin{equation}
d_{i k} = (1-\alpha_{k})d_{i(k-1)}  \label{eq:dj}
\end{equation} 
It may also happen that streamlines are occluded by multiple isosurfaces.
We can calculate the weights of segments of such streamlines by recursively applying equation (\ref{eq:dj}).
As a result, we calculate the information entropy by the following equation, supposing $1 \leq j \leq n_{j}$ and $d_{i_{0}} = d_{i}$:
\begin{equation}
E = - \frac{1}{ \log_{2} (1+m)} \sum^m_{j=0}\frac{d_{jn_{j}}}{l}\log_{2}\frac{d_{jn_{j}}}{l}\label{eq:enfinal}
\end{equation} 

\begin{figure}[ht]
  \begin{center}
   \includegraphics[width=50mm]{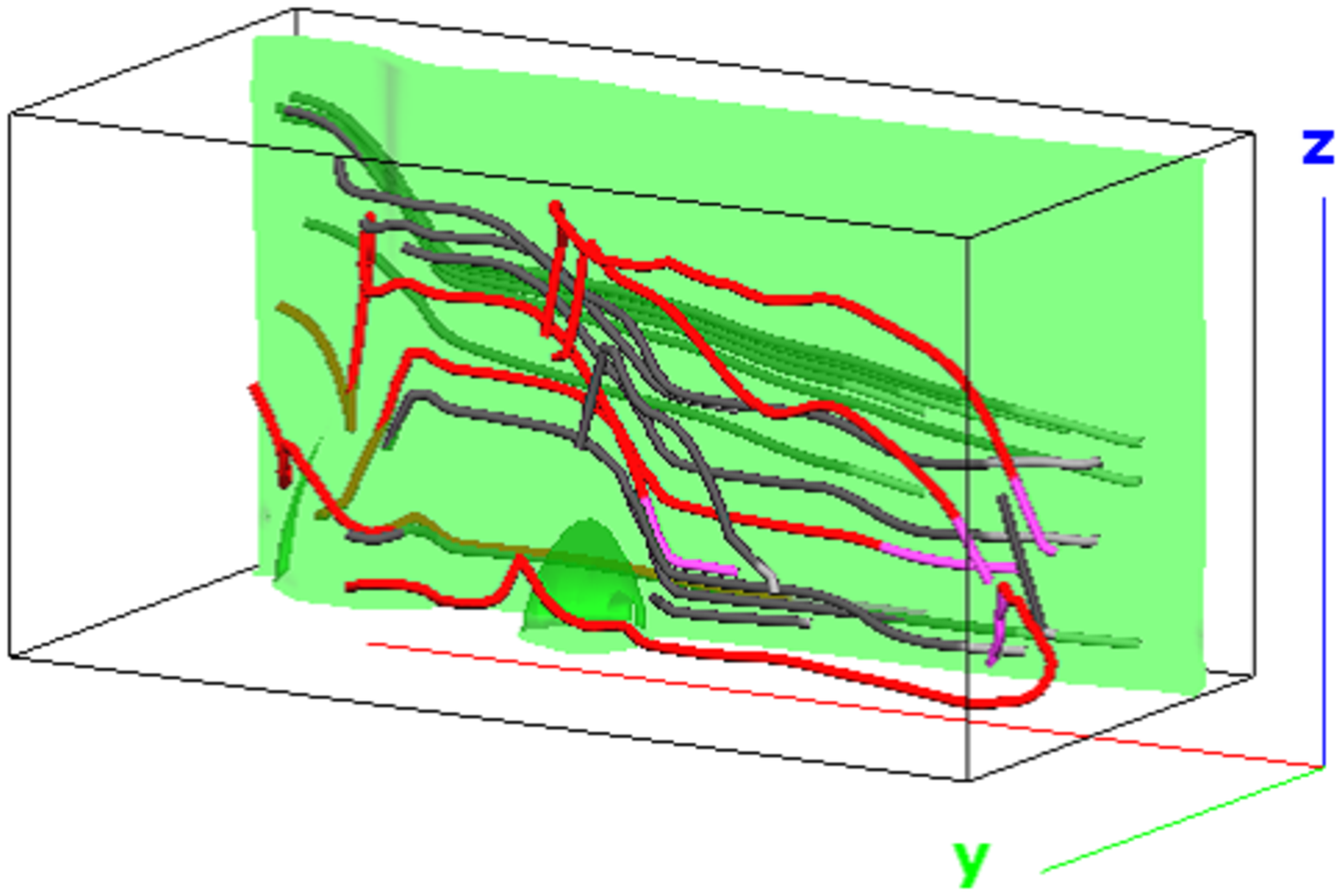}
   \includegraphics[width=50mm]{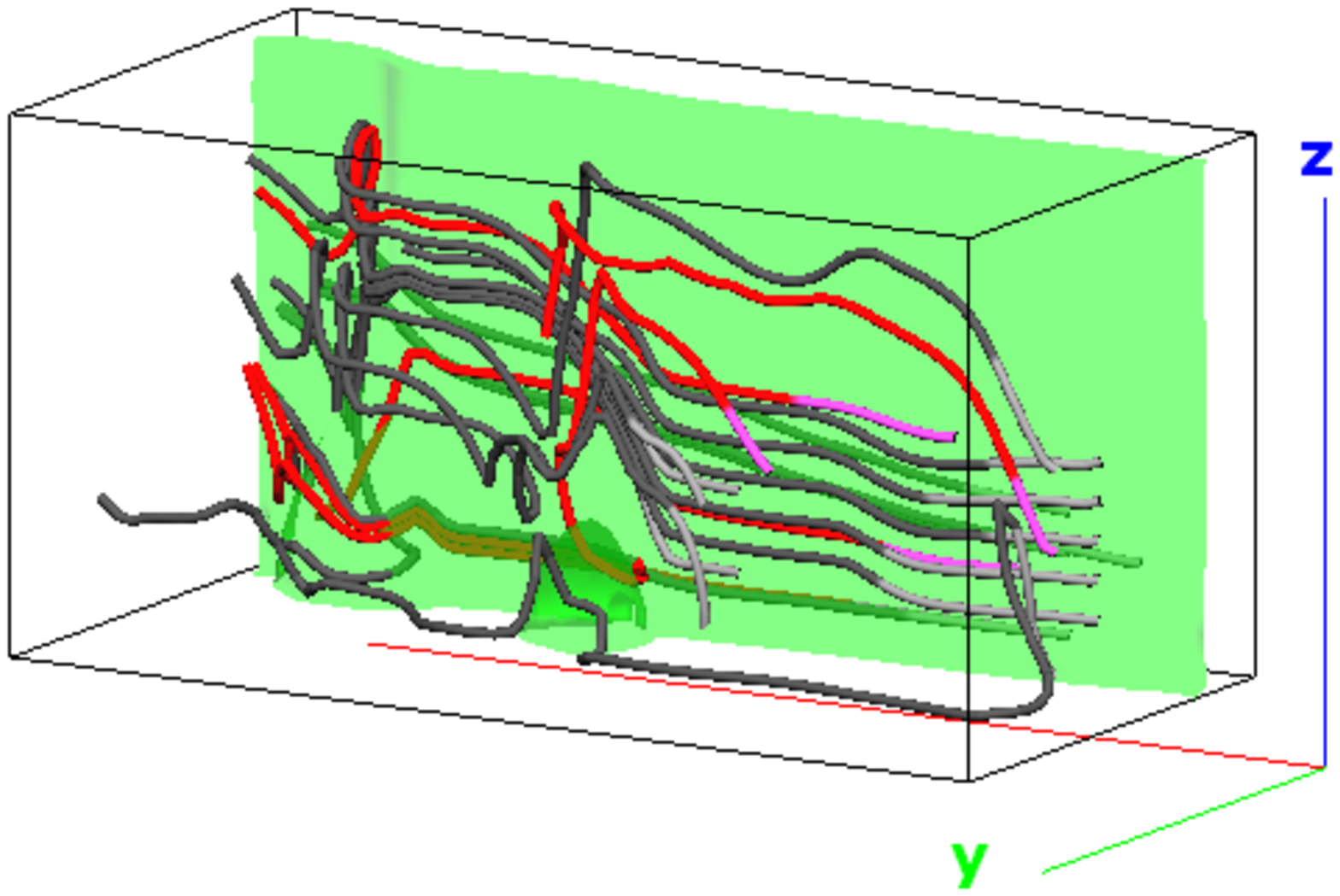}
  \end{center}
  \caption{Simultaneous display of isosurfaces and streamlines.
(Left) Without calculating weights based on occlusions by isosurfaces.
(Right) With calculating weights.}
  \label{fig:3-4}
\end{figure}

Figure \ref{fig:3-4} demonstrates the effectiveness of the weights based on occlusions of isosurfaces.
Figure \ref{fig:3-4}(Left) shows that many streamlines behind an isosurface is selected, while Figure \ref{fig:3-4}(Right) shows the presented technique successfully selects more visible streamlines.
As a result, our technique could automatically visualize a vortex in the left side of the input volume.

\subsubsection{Density of streamlines}
Our technique based on information entropy has another problem that large number of similar and close streamlines can be selected.
Accordingly, we determine if selected streamlines are sufficiently distant each other, so that we can avoid to densely select close streamlines.
Our current implementation exclusively selects one of the streamlines which intersect with the same cell, to avoid selecting too close streamlines.
However, this implementation does not always work well if the resolution of an input volume is finer comparing with resolution of a display.
We would like to implement an image-based streamline density control \cite{Li2007} to solve this problem.
Opacity control \cite{Gunther2013} is another solution to generate more comprehensive flow visualization for such situations.

\section{Example and evaluation}\label{sec:hyouka}

We implemented the presented technique with GNU C++ and OpenGL, and conducted subjective evaluation with the visualization results.
Participants of the evaluation were 12 university students majoring computer graphics and visualization.
We showed them pairs of visualization results and asked to select one of them.
Tables in the below sections show the number of participants who selected the corresponding visualization results.

\begin{figure}[ht]
\begin{center}
\includegraphics[width=100mm]{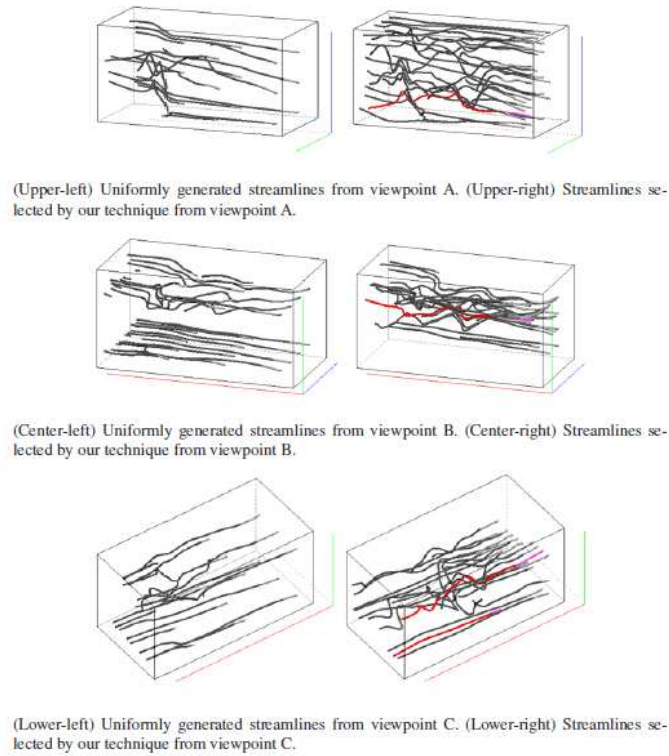}
\end{center}
\caption{\small Effectiveness of streamline selection. }
\label{fig:ABC_step1}
\end{figure}

\begin{figure}[ht]
\begin{center}
\includegraphics[width=100mm]{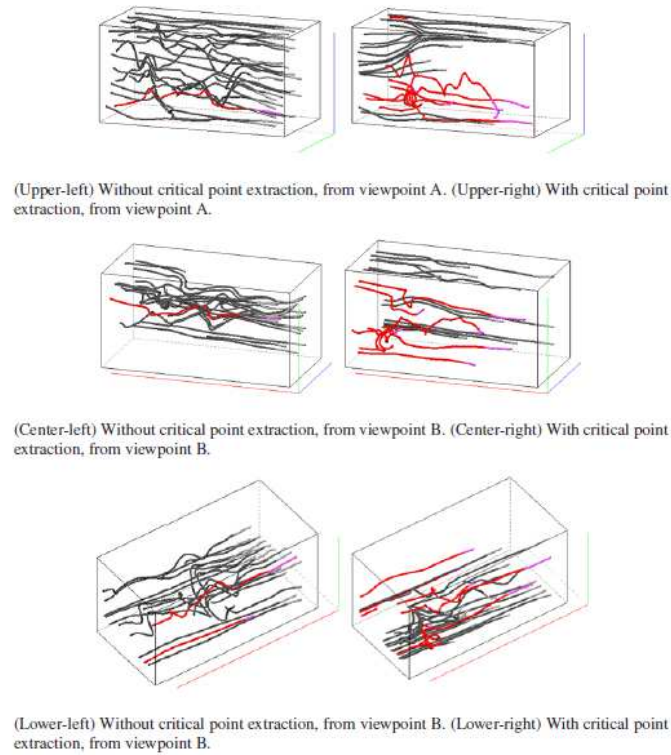}
\end{center}
\caption{\small Effectiveness of critical point extraction.}
\label{fig:ABC_step2}
\end{figure}

\begin{figure}[ht]
\begin{center}
\includegraphics[width=100mm]{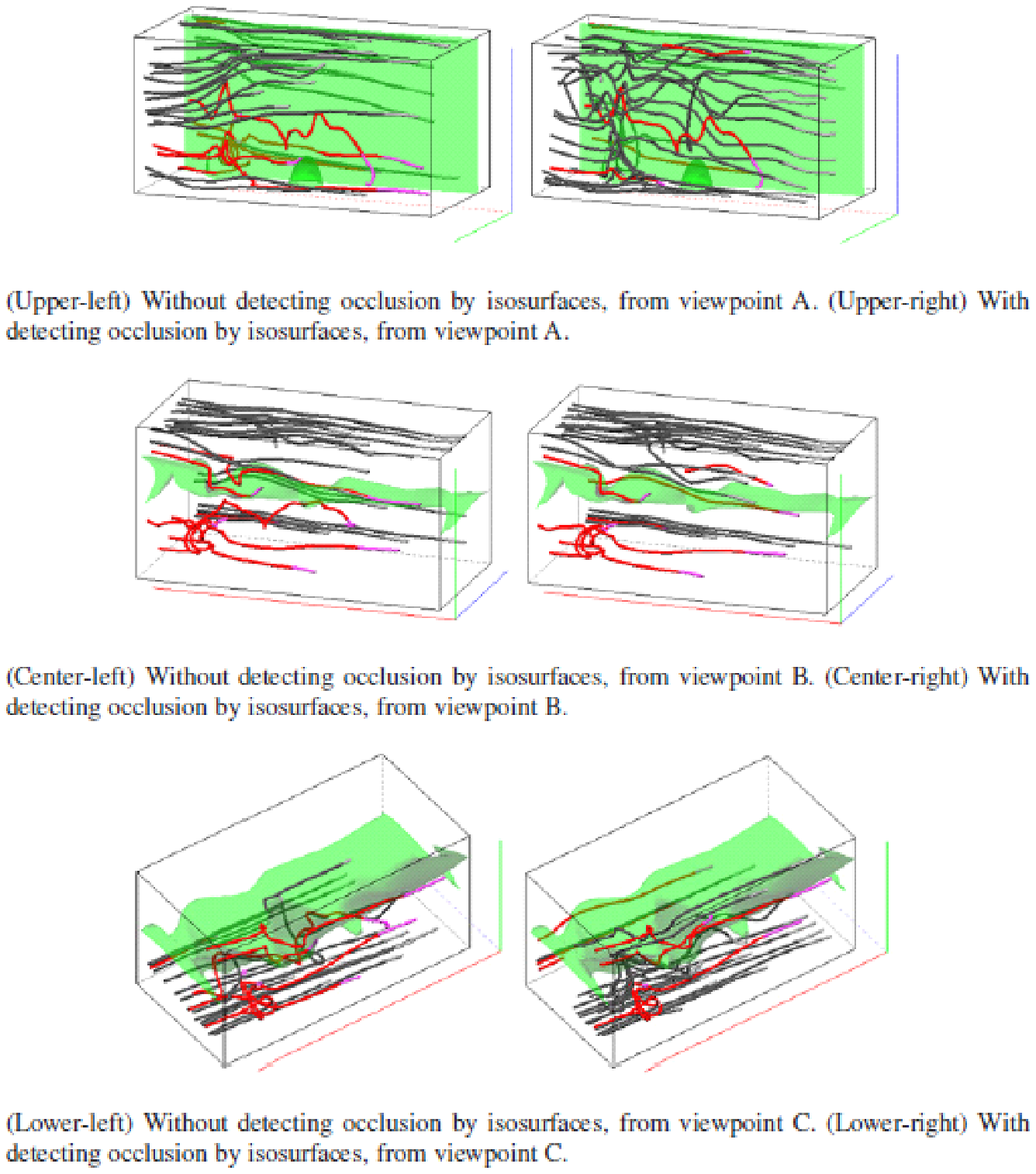}
\end{center}
\caption{\small Effectiveness of occlusion detection.}
\label{fig:ABC_step3}
\end{figure}

\subsection{Effectiveness of streamline selection}\label{sec:e-ent}

We generated visualization results with three viewpoints.
We used two implementations, where one uniformly generates streamlines, and the other displays streamlines automatically selected by the presented technique. 
Then, we asked participants to subjectively evaluate the results.
Figure \ref{fig:ABC_step1} shows the results.
Streamlines generated around critical points are drawn in red, while others are drawn in black.

We asked participants to select the visualization result which satisfies:
\begin{itemize}
\item {Red streamlines are more visible.}
\item {Black streamlines are less visible.}
\end{itemize}
Table \ref{tb:step1} shows the statistics of the evaluations of participants.

\begin{table}[ht]
\begin{center}
\caption{Subjective evaluation for effectiveness of streamline selection.}
\begin{tabular}{|c|c|c|}
\hline
 &\hfil Uniformly generated & \hfil Automatically selected \\
\hline
\hfil Viewpoint A & \hfil 1 & \hfil 11\\
\hline
\hfil Viewpoint B & \hfil 4 & \hfil 8\\
\hline
\hfil Viewpoint C & \hfil 2 & \hfil 10\\
\hline
\end{tabular}
\label{tb:step1}
\end{center}
\end{table}
The presented technique received preferable evaluations, especially with viewpoints A and C.
Following are typical comments for these results by the presented technique:

\begin{itemize}
\item {It is easier to understand the vector field since the presented technique selects longer streamlines.}
\item {The presented technique well represents local phenomena such as undulation of winds.}
\item {The presented technique shows larger number of red streamlines.}
\end{itemize}

On the other hands, four participants did not prefer the result by the presented technique with viewpoint B.  Their comments are as follows:
\begin{itemize}
\item {Visualization by the presented technique is too dense.}
\item {The result with uniformly selecting streamlines is simpler.}
\end{itemize}

\subsection{Effectiveness of critical point extraction}\label{sec:e-cpp}

We also asked participants to evaluate the results shown in Figure \ref{fig:ABC_step2}, as our previous questions introduced in Section \ref{sec:e-ent}.
This figure compares the visualization results whether critical points are extracted or not.

\begin{table}[ht]
\begin{center}
\caption{Subjective evaluation for effectiveness of critical point extraction.}
\begin{tabular}{|c|c|c|}
\hline
 &\hfil Without critical points & \hfil With critical points\\
\hline
\hfil Viewpoint A & \hfil 4 & \hfil 8\\
\hline
\hfil Viewpoint B & \hfil 4 & \hfil 8\\
\hline
\hfil Viewpoint C & \hfil 4 & \hfil 8\\
\hline
\end{tabular}
\label{tb:step2}
\end{center}
\end{table} 

Table \ref{tb:step2} shows the statistics of the evaluations of participants.
Extracting critical points and generating at least one streamline around each of the critical points, we could show more characteristic streamlines, and therefore received more preferable evaluations.
On the other hand, several participants mentioned that too many streamlines drawn in red might make difficult to understand the vector fields.

\subsection{Effectiveness of detection of occlusion by isosurfaces}

Finally, We asked participants to evaluate the results shown in Figure \ref{fig:ABC_step3}, as our previous questions introduced in Sections \ref{sec:e-ent} and \ref{sec:e-cpp}.
This figure compares the visualization results whether occlusion by isosurface is detected or not.

\begin{table}[ht]
\begin{center}
\caption{Subjective evaluation for effectiveness of occlusion detection.}
\begin{tabular}{|c|c|c|}
\hline
 &\hfil Without occlusion detection & \hfil With occlusion detection\\
\hline
\hfil Viewpoint A & \hfil 8 & \hfil 4\\
\hline
\hfil Viewpoint B & \hfil 6 & \hfil 6\\
\hline
\hfil Viewpoint C & \hfil 5 & \hfil 7\\
\hline
\end{tabular}
\label{tb:step3}
\end{center}
\end{table} 

Table \ref{tb:step3} shows the statistics of the evaluations of participants.
Our technique received relatively preferable evaluation while applying viewpoint C.
Following are major comments for the results with viewpoint C detecting occlusion by isosurfaces:
\begin{itemize}
\item {Red streamlines are not occluded by an isosurface.}
\item {Relationship between an isosurface and streamline is more comprehensive.}
\end{itemize}
On the other hand, our technique received relatively poor results with viewpoint A.
Following are major comments for the results with viewpoint A detecting occlusion by isosurfaces:
\begin{itemize}
\item {Visualization result got too complicated because too many streamlines are displayed in front of an isosurface.}
\item {Red streamlines are occluded by black streamlines rather than an isosurface.}
\end{itemize}

\section{Visualization of weather simulation data}\label{sec:jirei}

This section introduces a visualization of weather simulation dataset formed as a regular grid.
Number of cells are 80x40x80, where each grid-point has scalar values (air pressure and temperature) and a vector value (air flow).
Longitude and latitude are assigned to the X- and Y-axes, while height is assigned to the Z-axis.

First, we extracted critical points from a scalar (pressure) field. It contained 13 local maximum, 9 local minimum, and 25 saddle points.
This dataset simulates an air flow just before an extratropical cyclone is formed.

\begin{figure}[ht]
\centering
\includegraphics[width=55mm]{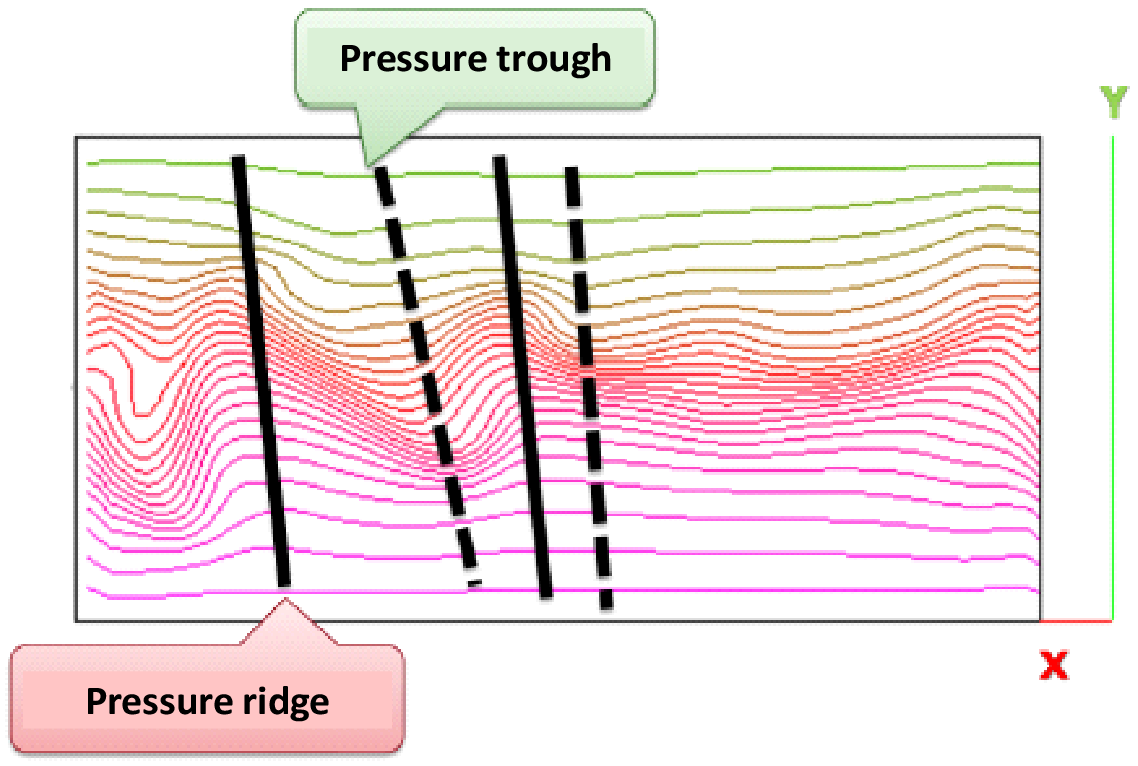}
\includegraphics[width=55mm]{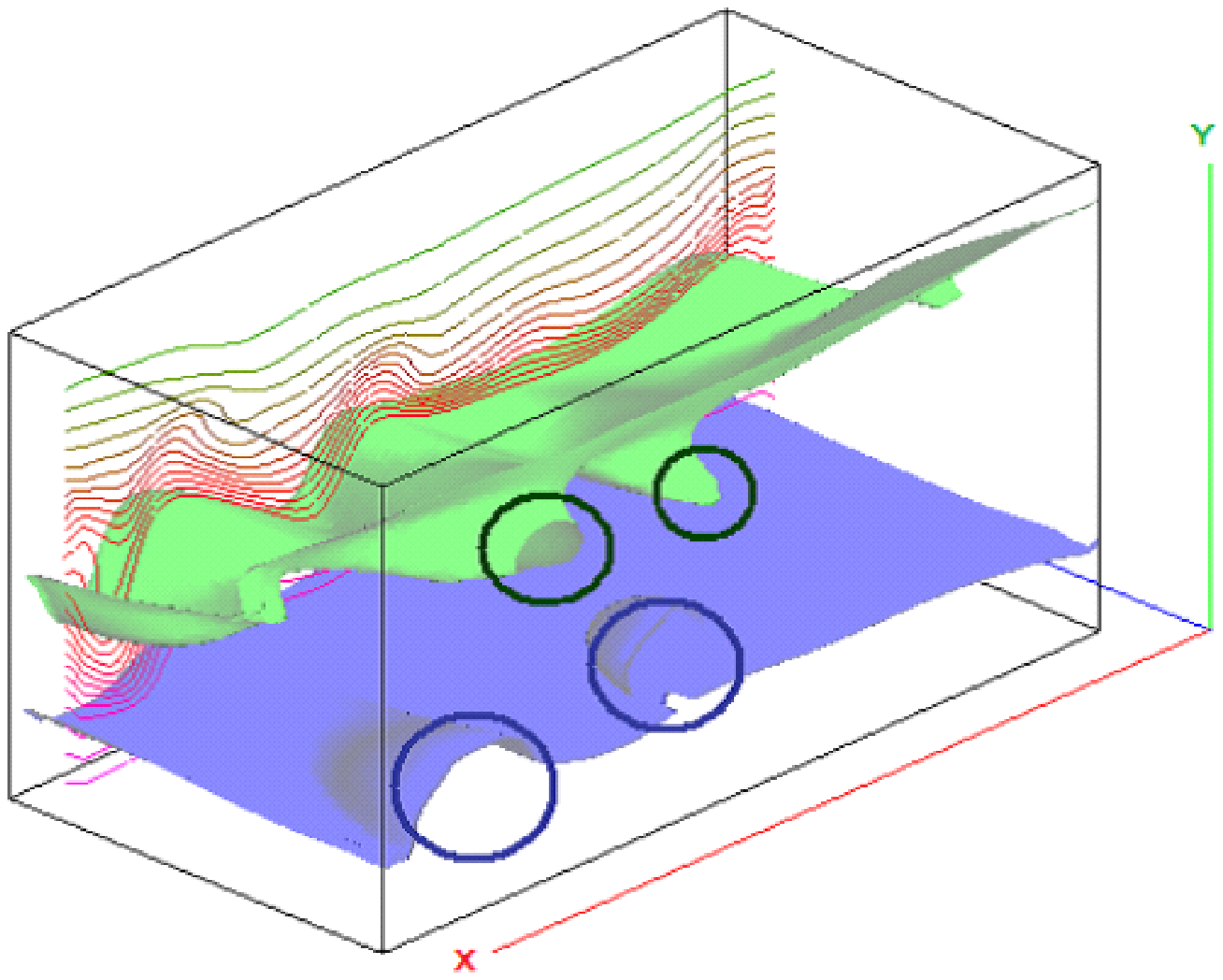}
\caption{\small
(Left) Air pressure on a XY-plane.
(Right) High and low pressure represented by isosurfaces.}
  \label{fig:jirei1}
\end{figure}

Figure \ref{fig:jirei1}(Left) shows iso-contour lines on a XY-plane.
Pressure ridge and trough are drawn as solid and dotted lines respectively.
In this result, pressure ridge is developed when westerlies swerves toward north, while pressure trough appears when westerlies swerves toward south. 

Figure \ref{fig:jirei1}(Right) shows a blue isosurface whose isovalue is slightly smaller than the pressure value at one of local maximum points, and a green isosurface whose isovalue is slightly larger than the pressure value at one of local minimum points.
High pressure regions appears where the blue isosurface curves, indicated by circles drawn in dark blue in Figure \ref{fig:jirei1}(Right).
Low pressure regions appears where the green isosurface curves, indicated by circles drawn in dark green in Figure \ref{fig:jirei1}(Right).

\begin{figure}[ht]
\centering
\includegraphics[width=100mm]{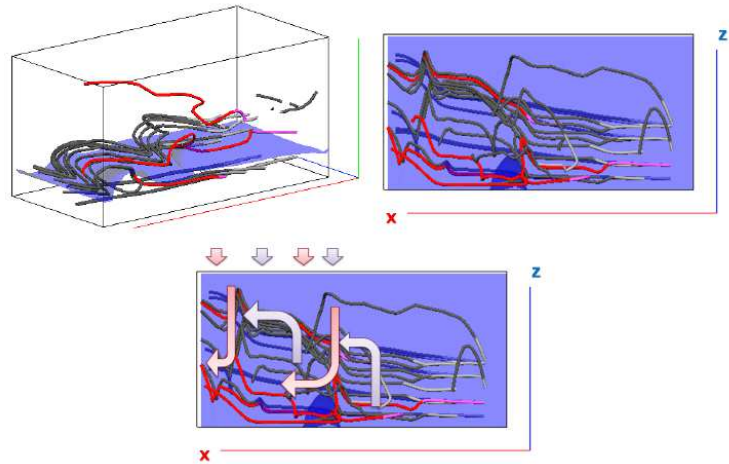}
\caption{\small
(Upper) Visualization of air flow around a high pressure region.
(Lower) Visualization of air flow around pressure ridge and trough.}
  \label{fig:jirei2}
\end{figure}

Figure \ref{fig:jirei2}(Upper) shows a visualization of air flow applying streamlines around a high pressure region.
This visualization displays streamlines around critical points in red, and the others in black.
Start points of the streamlines are drawn in subtle colors so that we can recognize these directions.
Left and right examples in Figure \ref{fig:jirei2}(Upper) display different sets of streamlines because our implementation recalculates information entropy when the viewpoint or view direction is switched.

Figure \ref{fig:jirei2}(Lower) illustrates pressure ridge and trough shown in Figure \ref{fig:jirei1}(Left) in addition to an isosurface and streamlines visualized in Figure \ref{fig:jirei2}(Upper).
Blue arrows depict pressure ridge, while pressure trough are illustrated by red arrows.
Downdraft is observed around pressure ridge, and updraft is observed around pressure trough.

High pressure regions are often developed when air temperature in the sky is much lower than one on the ground.
We can observe the relationship between high pressure regions and air temperature by switching the scalar field from pressure to temperature and then visualizing the temperature applying isosurfaces.

\section{Conclusion}

This paper presented an integrated visualization of scalar and vector fields applying an automatic streamline selection technique.
We select preferable streamlines from the following three aspects: geometric meaningfulness applying information entropy, distance from critical points, and occlusion by isosurfaces.
We conducted subjective evaluation with university student participants and concluded that information entropy and distance from critical points were especially effective factors of our technique.
We also introduced the effectiveness of our technique with a case study applying a weather simulation dataset.

The following are our potential future issues: \\
{\bf Improvement for volume dataset containing large number of critical points.}
The dataset we used as a case study just contained approximately 50 critical points in its scalar and vector fields.
It is difficult to deal with all the critical points if there are hundreds or even thousands in a dataset.
For example, it is not realistic to generate isosurfaces whose isovalues are close to values of all of thousands of local maximum or minimum points.
Or, it is also meaningless to generate streamlines around all of thousands of vortices.
We may need to select reasonable number of important critical points \cite{Takahashi2004} as a preprocess in such cases.\\
{\bf Extension for time-varying datasets.}
Our technique would select streamlines independently for each timestep, if we applied the current implementation to time-varying volume datasets.
This may cause flickering and prevent the understanding of time-varying air flow.
We would like to solve this problem by seamlessly selecting streamlines for each timestep of time-varying volume datasets.

\section*{Acknowledgement}

We appreciate Prof. Tetsuya Kawamura and Ms. Fumi Yasuda to provide us a weather simulation dataset.


\begin{thebibliography}{99}

\bibitem{Edmunds2012}
M. Edmunds, R. S. Laramee, G. Chen, N. Max, E. Zhang, C. Ware, Surface-Based Flow Visualization, {\it Computers \& Graphics}, 36(8), 974-990, 2012.
  
\bibitem{Furuya2008}
S. Furuya, T. Itoh, A Streamline Selection Technique for Integrated Scalar and Vector Visualization, {\it IEEE Visualization, Poster Session}, 2008. 

\bibitem{Gunther2013}
T. Gunther, C. Ross, H. Theisel,
Opacity Optimization for 3D Line Fields,
{\it ACM Transactions on Graphics (SIGGRAPH 2013)}, 32(4), 2013.

\bibitem{Hong1995}
L. Hong, X. Mao, A.Kaufman, Interactive Visualization of Mixed Scalar and Vector Fields, {\it IEEE Visualization}, 240-247, 1995.

\bibitem{Itoh1995}
T. Itoh, K. Koyamada, Automatic Isosurface Propagation by Using an Extrema Graph and Sorted Boundary Cell Lists, {\it IEEE Transactions on Visualization and Computer Graphics}, 1(4), 319-327, 1995.

\bibitem{Kindlmann1998}
G. Kindlmann, J. W. Durkin, Semi-automatic Generation of Transfer Functions for Direct Volume Rendering, {\it IEEE symposium on Volume visualization}, 79-86, 1998.

\bibitem{Koyamada1998}
K. Koyamada, T. Itoh, Seed Specification for Displaying a Streamline in an Irregular Volume, {\it Engineering with Computer}, 14, 73-80, 1998.

\bibitem{Laramee2004}
R S. Laramee, D. Weiskopf, J. Schneider, H. Hauser, Investigating Swirl and Tumble Flow with a Comparison of Visualization Techniques,  {\it IEEE Visualization}, 51-58, 2004.

\bibitem{Lee2011}
T.-Y. Lee, O. Mishchenko, H.-W. Shen, R. Crawfis,
View Point Evaluation and Streamline Filtering for Flow Visualization,
{\it IEEE Pacific Visualization Symposium}, 83-90, 2011.

\bibitem{Li2007}
L. Li, H.-W. Shen, Image Based Streamline Generation and Rendering, {\it IEEE Transactions on Visualization and Computer Graphics}, 630-640, 2007.

\bibitem{Li2008}
L. Li, H.-H. Hsieh, H.-W. Shen, Illustrative Streamline Placement and Visualization, {\it IEEE Pacific Visualization Symposium}, 79-86, 2008.

\bibitem{Marchesin2010}
S. Marchesin, C.-K. Chen, C. Ho, K.-L. Ma, View-Dependent Streamlines for 3D Vector Fields, {\it IEEE Transactions on Visualization and Computer Graphics}, 16(6), 1578-1586, 2010.

\bibitem{Mebarki2005}
A. Mebarki, P. Alliez, O. Devillers, Farthest Point Seeding for Efficient Placement of Streamlines, {\it IEEE Visualization}, 479-486, 2005.

\bibitem{Merhof2009}
D. Merhof, M. Meister, E. Bingol, C. Nimsky, G. Greiner,
Isosurface-Based Generation of Hulls Encompassing Neuronal Pathways,
{\it Stereotactic Functional Neurosurgery}, 87(1), 2009. 

\bibitem{Roger1993}
R. A. Crawfis, N. Max, Texture Splats for 3D Scalar and Vector Field Visualization, {\it IEEE Visualization}, 261-266, 1993.

\bibitem{Sch2012}
P. Schaefer, M. Gampert, N. Peters,
Streamline Segment Topology in the Vicinity of Stagnation Points in Turbulent Flows,
{\it 15th International Symposium on Flow Visualization}, 2012.

\bibitem{Takahashi2004}
S. Takahashi, Y. Takeshima, I. Fujishiro, Topological Volume Skeletonization and Its Application to Transfer Function Design, {\it Graphical Models}, Vol. 66, No. 1, 24-49, 2004.

\bibitem{Takahashi2005}
S. Takahashi, I. Fujishiro, Y. Takeshima, T. Nishita, A Feature-Driven Approach to Locating Optimal Viewpoints for Volume Visualization, {\it IEEE Visualization}, 495-502, 2005.

\bibitem{Tao2013}
J. Tao, J. Ma, C. Wang, C.-K. Shene,
A Unified Approach to Streamline Selection and Viewpoint Selection for 3D Flow Visualization,
{\it IEEE Transactions on Visualization and Computer Graphics}, 19(3), 393-406, 2013.


\bibitem{Xu2010}
L. Xu, T.-Y. Lee, H.-W. Shen, An Information Theoretic Framework for Flow Visualization, IEEE Transactions on Visualization and Computer Graphics, 16(6), 1216-1224, 2010.

\bibitem{Ye2005}
X. Ye, D. Kao, A. Pang, Strategy for Seeding 3D Streamlines, {\it IEEE Visualization}, 471-478, 2005.

\end{thebibliography}
\end{document}